\documentclass[letterpaper,twoside,english]{article}
\usepackage[T1]{fontenc}
\usepackage[latin1]{inputenc}

\makeatletter

\usepackage{setspace}

\usepackage{latexsym,times}
\usepackage[intlimits]{amsmath}
\usepackage{amsfonts,amssymb}
\DeclareSymbolFontAlphabet{\mathbb}{AMSb}
\usepackage[mathscr]{eucal}
\usepackage{fancyhdr}
\usepackage[notoday]{rcsinfo}
\usepackage{changebar} %\nochangebars
\usepackage[textstyle,cdot,squaren,thickspace,thickqspace]{SIunits}
\usepackage{fancyvrb}
\usepackage{url}
\usepackage{xspace}
\usepackage[usenames,dvips]{color}
\usepackage[dvips]{graphicx}
\DeclareGraphicsExtensions{.eps}   % extension for included graphics
\graphicspath{
  {./}
}
\usepackage{psfrag}
\usepackage[ps2pdf]{thumbpdf}              % thumbnails for ps2pdf
\usepackage[%
bookmarks=true,%                 % generate bookmarks ...
bookmarksnumbered=true,%         % ... with numbers
hypertexnames=false,%            % needed for correct links to figures !!!
pdfborder={0 0 0},%  % border-width of frames (multiplied with 0.009 by ps2pdf)
ps2pdf]{hyperref}%
\hypersetup{
  pdfauthor   = {can ozan tan <tanc@cns.bu.edu>},
  pdftitle    = {},
  pdfsubject  = {},
  pdfkeywords = {},
  pdfcreator  = {LaTeX with hyperref package},
  pdfproducer = {dvips and ps2pdf}
}
\usepackage[letterpaper,top=1in,bottom=1in,hmargin={1in,1in}]{geometry}
\numberwithin{equation}{section}
\DeclareRobustCommand\em
{\@nomath\em \ifdim \fontdimen\@ne\font >\z@
  \upshape \else \slshape \fi}

\usepackage{babel}

\usepackage{babel}

\usepackage{babel}
\makeatother
\begin{document}

\title{ECOSYSTEMS IN THE MIND: FUZZY COGNITIVE MAPS OF THE KIZILIRMAK DELTA
WETLANDS IN TURKEY}

\author{Uygar \"{O}zesmi}
\date{}
\maketitle
\begin{spacing}{0.5}
\begin{center}Department of Environmental Engineering\end{center}

\begin{center}Erciyes University, 38039 Kayseri, Turkey\end{center}

\begin{center}uygar@ozesmi.org \end{center}
\end{spacing}

\begin{abstract}
Sustainability of ecosystems and ecosystem management are increasingly
accepted societal goals. However projects of non-governmental conservation
organizations (NGOs) and governments also are prone to becoming impractical
impositions on local people. Can conservation programs be improved
by incorporating local people's understanding of ecosystems? The Kizilirmak
Delta is one of Turkey's most important wetland complexes with its
rich biodiversity and critical habitat for globally endangered bird
species. It is also one of the most productive agricultural deltas
in Turkey. We have drafted, together with stakeholders, 31 cognitive
models of the social and ecological system. These models were converted
to adjacency matrices, analyzed using graph theoretical methods, and
augmented into social cognitive maps. Causal models were run based
on neural network computational methods. \char`\"{}What-if\char`\"{}
scenarios were run to determine the trajectory of the ecosystem based
on the ecosystem models defined by stakeholders. Villagers had significantly
larger numbers of variables, more complex maps, a broader understanding
of all the variables that affect the Kizilirmak Delta, and mentioned
more variables that control the ecosystem than did NGO and government
officials. Villagers had developed a large capacity to adapt to changing
ecological and social conditions. They actively changed and challenged
these conditions through the political process. Villagers were faced
with many important forcing functions that they could not control.
Most of the variables defined by villagers were related to agriculture
and animal husbandry. Conservation policies and ecosystem management
must encompass larger environmental issues that villagers raise as
much as biodiversity and villagers' cognitive maps must be reconciled
with that of NGOs and government officials. Cognitive maps can serve
as a basis for discussion when policies and management options are
formulated. A villager-centered cognitive mapping approach is not
only necessary because villagers resist conservation projects, or
because top down projects that do not take local knowledge systems
into account fail, but because it is the ethical and responsible way
of doing ecosystem management.

\end{abstract}

\section*{INTRODUCTION}

The Kizilirmak Delta is one of Turkey's most important wetland complexes
with its rich biodiversity and critical habitat for globally endangered
bird species. It is also one of the most productive agricultural deltas
in Turkey. Because of these qualities, the delta has been declared
a Ramsar Site, a wetland of international importance by the Ministry
of Environment (Resmi-Gazete 1998). The Delta is located in the central
Black Sea region of Turkey (41°30 to 41°45' N, 35°43' to 36°08' E),
and covers an area of 50,000 ha that includes 15,000 ha of freshwater
marshes and swamps, coastal lakes, and lagoons (Fig. 1). More than
310 bird species, or 75\% of all known bird species in Turkey, use
the delta for breeding, wintering, and migration. It hosts several
globally threatened species including the Dalmatian Pelican (\textit{Pelecanus
crispus}), White-headed Duck (\textit{Oxyura leucocephala}), Red-breasted
Goose (Branta ruficollis), Ferruginous Duck (\textit{Aythya nyroca})
and Imperial Eagle (\textit{Aquila heliaca}) (Hustings and Dijk 1993).
Limited studies report 10 species of mammals, 8 species of reptiles,
8 species of fish, and 18 species of invertebrates in the delta (Hustings
and Dijk 1993; Ozesmi and Karul 1990).

\begin{figure}[ht]
\begin{center}
\includegraphics[width=5in]{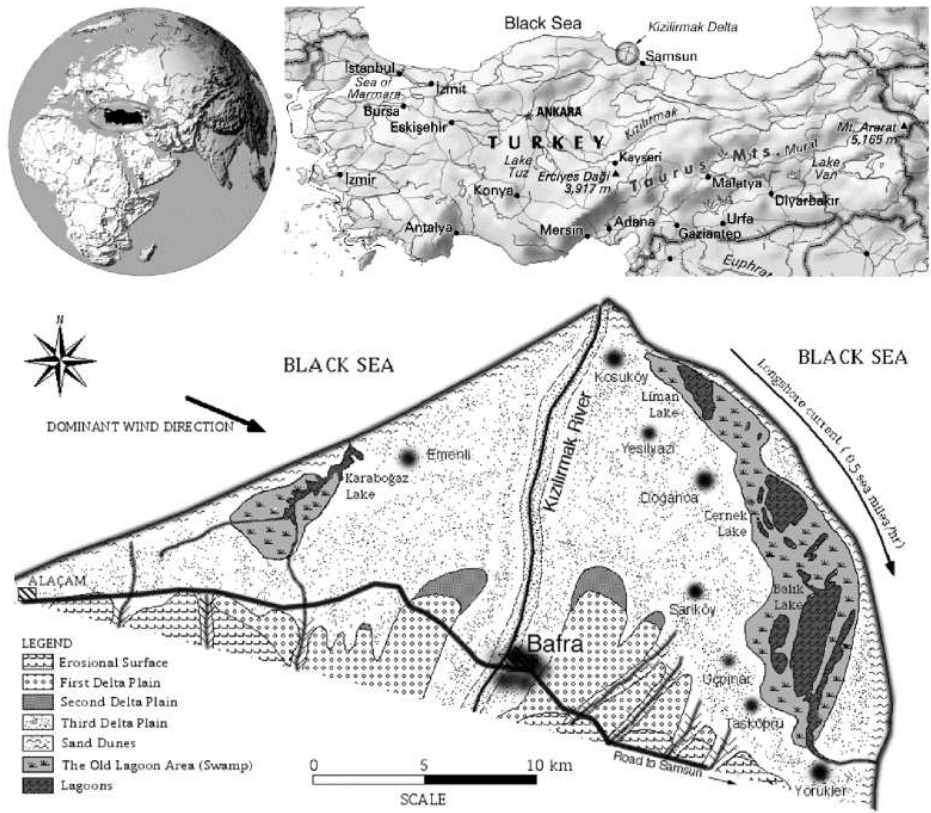}
\caption{The location of the Kizilirmak Delta at the Black Sea coast of Turkey. The delta is constituted mainly of the third delta plain that has formed in the last 10,000 years. The shape and formation of lagoon barriers are controlled by sediment influx from the rivers and longshore currents (Akkan 1970; Ozesmi 1992). Behind lagoon barriers, 15,000 ha of wetland and lagoon habitat has formed.}
\end{center}
\end{figure}

Human populations have settled around wetland ecosystems since prehistory
(Coles and Coles 1992). People have inhabited the Kizilirmak Delta
at least since the Chalcolithic (Alkim et al. 1988) . They have used
many products extracted from the wetlands including fish, waterfowl,
thatch, commercial harvest of aquatic vegetation, hay, fodder, timber,
and more recently rice agriculture and recreation (Alkim et al. 1999;
Ozesmi, 2003; Tekkaya and Payne 1988). Today all of the first and
second delta plains and most of the third delta plain is dominated
by agriculture, crisscrossed by drainage and irrigation canals and
roads. Settlements are dispersed with small village centers at crossroads
typically including a mosque, school, health center, tea-house and
grocery store.

It is not uncommon for a heated discussion to break out in the tea-houses
or homes around the wetlands on the future of the Kizilirmak Delta
and the livelihood of the villagers. A conflict exists over the future
of the delta between the stakeholders; the villagers, vacation home
owners, government entities and the NGOs. The Department of National
Parks has wanted to declare the area a national park since the 1970s.
After learning about the plans for a national park, a group of village
elders visited the first bird sanctuary, Manyas Lake, a national park
in western Turkey in 1985. The villagers around Manyas Lake told the
delegates how their livelihood practices were disrupted by the park.
When delegates reported to the village what they had been told, villagers
reached a consensus opinion against the national park. The aversion
to the park increased when wealthier villagers who were engaged in
illegal real estate deals in ecologically sensitive areas came into
conflict with NGOs that were working to conserve the Kizilirmak Delta
ecosystem. As in many other ecologically sensitive areas an absolute
protection strategy failed and conflict escalated among stakeholders.

But, there is also a consensus among stakeholders that the wetland
ecosystem is rapidly degrading. Alternative strategies are urgently
needed for the conservation of this unique ecosystem. Therefore, the
implicit normative goal of this paper is the conservation of the Kizilirmak
Delta. However, for this goal, I do not reason based on consequentialist
ethics, but argue that any viable conservation strategy in the future
has to involve different stakeholders and must integrate in the process
local peoples concerns and understanding of ecosystem with that of
government and NGO officials.

In this study I attempt to identify and reconcile the views of different
the stakeholders on how the Kizilirmak Delta ecosystem functions.
Stakeholders are encouraged to define the most important variables,
and the cause and effect processes in the delta. Sustainable resource
use for the conservation of the wetland ecosystem will not be adopted
by the villagers if it does not originate from their own perceptions
of the system. I use individual and social fuzzy cognitive maps of
stakeholder groups to analyze the ways in which the perceptions of
these groups are both different and similar. These different and shared
knowledges are drawn upon to develop viable conservation strategies.
I run simulations of different conservation strategies based on the
social cognitive maps of the stakeholders to provide policy options.

The information obtained from this study should be valuable for developing
policies and strategies to help ensure the sustainability of the Kizilirmak
Delta wetland ecosystem as well as the livelihoods of the people living
around them. The approach and results may provide insights to the
conservation of similar ecosystems where a conflict exists among different
stakeholders.

\section*{METHODS}

\subsection*{Introduction to Fuzzy Cognitive Maps}

Fuzzy cognitive maps have their roots in graph theory, first formulated
by Euler in 1736 (Biggs et al. 1976). Since then graph theory has
been developed extensively by mathematicians, but Harary et al. (1965)
first presented the theory of directed graphs (digraphs) for investigators
studying the structural properties of the empirical world. This empirical
world in our case is the Kizilirmak Delta wetland ecosystem. Anthropologists
have used signed digraphs to represent different social structures
in human society (Hage and Harary 1983). Axelrod (1976) transferred
signed digraphs from an empirical realm as understood by the anthropologist
to the assertions of the informants. For this representation Axelrod
(1976) coined the term cognitive map (first used by Tolman (1948))
that graphed causal relationships among variables as defined and described
by people. Numerous studies have used cognitive mapping to look at
decision making and conceptions of complex social systems (Axelrod
1976; Bauer 1975; Bougon et al. 1977; Brown 1992; Carley and Palmquist
1992; Cossette and Audet 1992; Hart 1977; Klein and Cooper 1982; Malone
1975; Montazemi and Conrath 1986; Nakamura et al. 1982; Rappaport
1979; Roberts 1973). These studies opened the way for using cognitive
maps to represent local knowledge systems as told by informants. In
this study the cognitive maps were the causal descriptions of the
Kizilirmak Delta Ecosystem as drawn by villagers, NGO and government
officials. The main assumption of this approach is that individuals
have cognitive models that are internal representations of a partially
observed world (Bauer 1975). This assumption has been verified quite
extensively by the works of cognitive psychologists (Anderson 1983),
and physiologists, anthropologists and sociologist have studied how
humans construct these cognitive maps (Bateson 1972; Bateson 1979;
Hannigan 1995; Jerison 1973). There have been many studies that have
used a cognitive mapping approach. Klein and Cooper (1982, p. 64).
succinctly summarize the new perspective that this type of research
offers: \char`\"{}Traditionally, scientists ... have been concerned
with the real, objective world, where phenomena can be observed and
measured. However, human decision processes never take place in this
objective world ... Human decision processes always take place within
the subjective world of the individual decision-maker. Cognitive mapping
offers a window on this subjective world.\char`\"{}

\subsection*{Fieldwork}

As a starting point before going out into the field, I constructed
my own cognitive map of the Kizilirmak Delta ecosystem giving real
numbers to edges (causal connections) in the interval {[}-1,1{]}.
In doing so I introspectively questioned my own assumptions and understanding
of the ecosystem as they had evolved since I first visited the delta
in 1989. This enabled me to be aware of my model and not be unconsciously
influenced by it, when I was drawing cognitive maps with the informants.
I recommend this exercise to be done before going to the field to
any future research on cognitive mapping.

In the summers of 1997 and 1998, I spent a total of six months in
Turkey doing research for developing sustainable resource use strategies
for the Kizilirmak Delta. I lived in villages and fishing huts located
around the wetlands of the Kizilirmak Delta. I interviewed villagers
in their spare time, as well as local NGO and government officials
in Bafra and Samsun and central NGO and government officials in Istanbul
and Ankara. I conducted in-depth interviews (Denzin and Lincoln 1994)
and drew cognitive maps with villagers and officials (Appendix I).
The structure of the interviews was based on model primitives (concepts
and statements) and used the first two steps of textual analysis for
drawing cognitive maps as described by Carley and Palmquist (1992).
The interview started by asking an open-ended question: \char`\"{}What
are the most important variables (concepts) in the Kizilirmak Delta\char`\"{}.
If I was asked to further clarify the question I said: \char`\"{}Variables
that make the Kizilirmak Delta, variables related to your experiences,
livelihood and the sustainable development of the delta, plain, which
ever is relevant to you.\char`\"{} After the informant listed all
variables and had nothing more to add, the variables were put on a
sheet of paper as circled units and the connections (edges, statements)
among variables were explored with informants (Appendix I). The interviewing
method I used is more direct than textual analysis, coding from a
printed text, and allows for cognitive maps directly drawn on paper
by the interviewees . The informants drew lines with arrows and +/-
signs from variables they selected to another variable showing causal
decrease or increase. They were asked to give the strength of the
causal relationship as a real number in {[}-1,1{]} or say whether
it was strong, medium, or low. The informants were encouraged to add
variables to their initial list as they seemed appropriate at this
stage. The informant was invited to describe and clarify stated relationships
and notes were taken. Interviews lasted an average of one and a half
hours in the range of 45 minutes to four hours. Each interview was
terminated when informants were satisfied with their cognitive map
and had nothing more to add. In this study I found cognitive mapping
to be especially useful because of a previous aversion to questionnaire
techniques encountered in the region. The openness and transparency
of cognitive maps makes them ethically more defensible compared to
other methods. I also preferred cognitive mapping because active drawing
with informants facilitates a broader and holistic discussion on the
delta ecosystem, conservation and livelihood practices providing insights
to reasoning, beliefs, and values. Brown (1992) presents further pro
et contra aspects of cognitive mapping technique compared to other
potential methods.

The final product was a cognitive map with the most important variables
and the causal relationships that were perceived between these variables
for each informant. A total of 31 cognitive maps were used in the
analysis. Of these 31 maps, 15 were drawn by inhabitants of the Kizilirmak
Delta coming from 5 villages, 2 were drawn by vacation home owners,
7 were drawn by local and national NGO officials, 7 were from government
officials. Of the 7 cognitive maps from government officials, 3 were
drawn during the interview and 4 were extracted from the conversational
interview text by coding, based on the concepts that they have stated
at the start of the interview (Carley and Palmquist 1992; Wrightson
1976).

\subsubsection*{Neural Network Computation}

The cognitive maps were then transformed according to graph theory
into adjacency matrices in the form $A(D)=[a_{ij}]$ (Harary et al.
1965) where the variables $\nu_{i}$ (e.g. pollution) were listed
on the vertical axis and $\nu_{j}$ (e.g. fish population) on the
horizontal axis forming a square matrix. Matrix entries were made
based on the cognitive map. For example -1 was entered for $a_{ij}$
if there was a causal decrease from $\nu_{i}$ to $\nu_{j}$ (eg.
pollution decreased fish populations) (Fig. 2). The adjacency matrices
of the cognitive maps where then treated as Fuzzy Cognitive Maps (FCM).

\begin{figure}[ht]
\begin{center}
\includegraphics[width=4in]{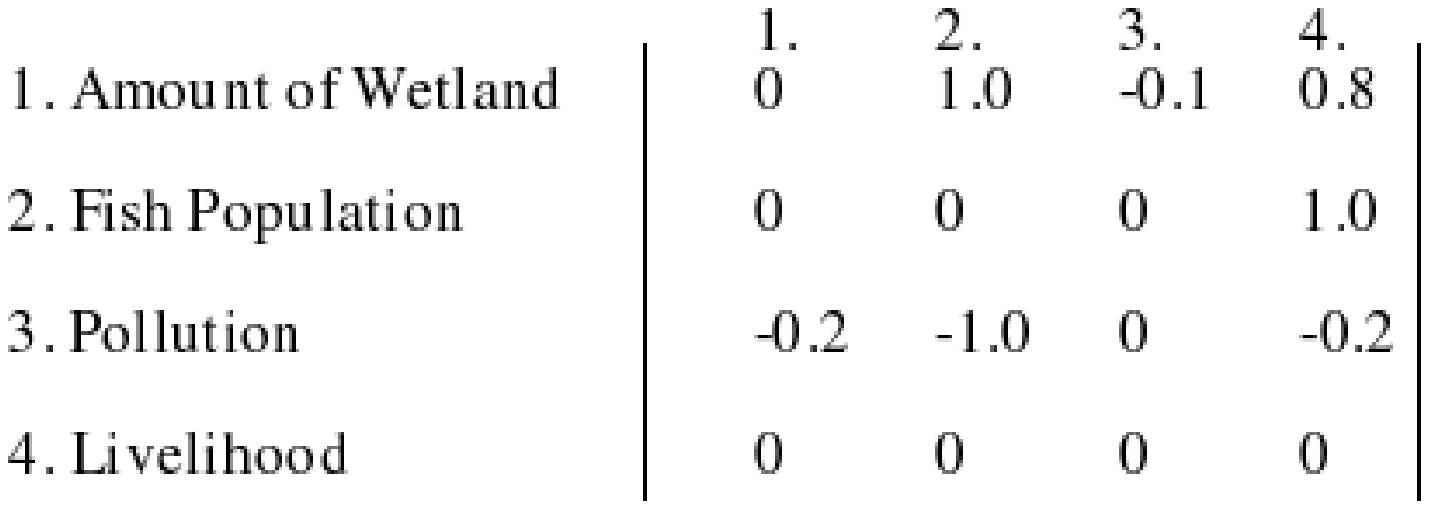}
\caption{A simplified illustration of an adjacency matrix, where pollution ($\nu_3$) causes a decrease of -1 ($a_{32}$) in fish population ($\nu_2$).}
\end{center}
\end{figure}

FCMs are useful in symbolic representations of semantic networks and
they also provide a method for modelling cognitive and empirical cause-effect
processes. Their advantage lies in that they allow feedback processes
(Kosko 1992b). Anyone can freely draw causal pictures of problems
or systems of any kind. For example, Taber (1991; 1987) used FCMs
to model physiology of food appetite and political developments. Styblinski
and Meyer (1988) used them to analyze electrical circuits. Gotoh and
Murakami (1989 as cited in Kosko 1992) used FCMs to model industrial
plant control. Dickerson and Kosko (1994) used them to model marine
food web interactions. Craiger et al. (1996) used them to simulate
organizational behavior and job satisfaction. Radomski and Goeman
(1996) used FCM to suggest ways to improve decision making in sport-fisheries
management. Schneider et al. (1998) used United Nations data on 106
countries to automatically build an economic/demographic model of
world nations from statistical data.

Kosko (1986) coined the term FCM when he modified Axelrod's cognitive
maps by applying fuzzy causal functions with real numbers in {[}-1,
1{]} to connections (also called lines, edges, synapses). Auto-associative
neural network architecture (Reimann 1998) is the basis of FCM. Neural
networks (also called connectionist architectures and parallel distributed
processing systems) are groups of simple, highly interconnected processing
units which act together to perform computable functions (Fig. 3).
FCMs allow 'causal inferences to be made as feedback associative memory
recollections' (Kosko 1987). The inference was accomplished in the
cognitive maps representing the Kizilirmak Delta Ecosystem by a neural
network computational method in which a vector of initial states of
variables ($I^{n}$) was multiplied with the adjacency matrix $A$
of the cognitive map. The matrix values are of variable strength,
that are represented by real numbers. The lines carry the input from
one variable (what is called a point, node, or unit) to another activating
the unit. The contribution of one connection to the unit is the product
of the activity on the line and the value of the connection strength.
The total input to the unit is the sum of all the individual products
(Fig. 3). Lines can be positive or negative. Positive connections
add to the activity total, while negative connections subtract from
it. The output of the unit is a function of the total input. In this
inference method usually a threshold function or a transformation
by a bounded monotonic increasing function is applied to the result
of the matrix multiplication, $I^{n}\times A$, at each simulation
time step (Kosko 1987; Kosko 1992b). Commonly used activation functions
are logistic, linear threshold or step functions. I used a logistic
function $1/(1+e^{-X})$to transform the results into the interval
{[}0,1{]}. This non-negative transformation allows for a better understanding
and representation of activation levels of variables and enables a
qualitative comparison among the causal output of variables. The resulting
transformed vector then was signaled through the adjacency matrix
and transformed repeatedly until the system converged to a fixed point
in less than 30 simulation time steps. Theoretically it could have
also settled into a limit cycle, or chaotic attractor (Dickerson and
Kosko 1994).

Using this neural network computational method it is possible to run
'what-if' scenarios using the initial state vector. Specific variables
related to a scenario can be 'clamped' in the vector by setting these
variables at a desired value at each simulation step (Kosko 1988).
The final result indicates where the system would end up given initial
variable states or simulated policy options.
\vspace{0.5in}
\begin{figure}[ht]
\begin{center}
\includegraphics[width=2in]{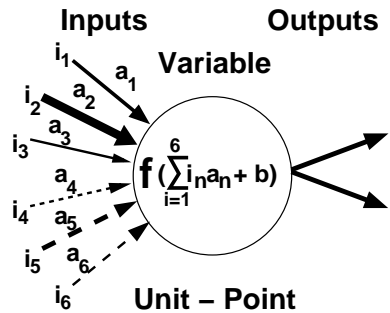}
\caption{A simulated unit representing a variable in the cognitive ecosystem model. Black arrows represent positive signed and gray ones negative signed connections. All inputs are multiplied by weights of connections and then summed. The summation is input into a monotonic increasing function. The result is then output to other connected units.}
\end{center}
\end{figure}

\subsubsection*{Social Cognitive Maps}

Laszlo et al. (1996) state that a set of cognitive maps of individuals
form a social cognitive map that behaves differently than a simple
addition or overlay of individual cognitive maps. Once a social cognitive
map forms it takes a life on its own through social processes, recursively
affecting individuals who act not only based on their own cognitive
maps but are affected by the social cognitive maps. As Eden (1988,
p.7) states: \char`\"{}An aggregated model constructed by combining
each of the individual cognitive maps produces a \char`\"{}team
map\char`\"{} that is no longer a representation of the cognition
/thinking of anyone and does not belong to anyone.\char`\"{} Social
knowledge is defined as shared knowledge (Carley 1988) and is usually
assumed to be obvious and may not be stated in cognitive maps readily.
Therefore, the interviewer must have at least a rudimentary awareness
of the study groups assumptions and beliefs and ask appropriate questions
to elicit such social knowledge.

Based on Laszlo et al. (1996), I additively superimposed individual
cognitive maps of informants that resulted in fuzzy cognitive maps
(Kosko 1987; Kosko 1992a; Kosko 1992b) to form a social cognitive
map. The resulting augmented social cognitive map shows different
properties and results then an addition of the neural network computation
results of individual cognitive maps. In the augmentation, while conflicting
connections with opposite signs cancelled each other out, agreement
reinforced causal relationships, forming a consensus social cognitive
map based on equal representation. Zhang and Chen (1988) point out
that the differing directions might be a result of different logical
structure and suggest the use of a negative- positive-neutral calculus
to compute compound values for augmented maps. In my study, the augmentation
of maps was based on the equivalence properties of fuzzy causal relationships
(FCR) among variables (Kim and Lee 1998). In augmented cognitive maps
informant contributions can be scaled either by multiplying their
adjacency matrix by a subjective weight or by assigning weights based
on the degree of concurrence (Taber 1991; Taber and Siegel 1987).
Schneider et al. (1998) point that weights based on concurrence might
represent conservative maps at the expense of people thinking differently.
Therefore, in this study an egalitarian assumption was made that all
individual maps were equally valid and a weight of one has been assigned
to each individual cognitive map. The social cognitive maps in this
study represent only one point in time. One has to consider that a
social cognitive map is not static in time and evolves as the community
itself transforms. As Carley (1997) points out social cognitive maps
are \char`\"{}lossy consensus\char`\"{} that dynamically change
over time loosing parts by members leaving a community and others
joining in. Community members might not have developed a shared language
during their time together. Although they might have overlapping common
beliefs about certain aspects of the systems, the partial shared beliefs
do not guarantee that they can draw a social cognitive map together
through a structured protocol. Power within group members can create
an unequal environment of expression and sometimes conflict when drawing
cognitive maps (Langfield-Smith 1992). Malone (1975) in a group exercise
with eight students could also not come to a consensus and had to
resort to voting in determining map connections. Drawing social cognitive
maps in groups was not attempted in this study and all participants
had equal representation in the final social cognitive maps.

Informants may draw cognitive maps that emphasize different parts
of a system based on their experiences. This need not imply that these
maps are wrong or less representative. To the contrary, the addition
of these cognitive maps might yield a better representation of the
system (Eden et al. 1979). Roberts (1973) states that the larger groups
of experts yield more accurate and reliable information, and Nakamura
et al. (1982) found that joining of cognitive maps generated useful
information that was not captured by individual maps. In statistical
terms, large numbers of independent and identically distributed (i.i.d.)
observations will tend to produce stable edge values (Kosko 1988),
and exponentially diminishing number of new variables. The slow accumulation
of variables might be due to a limited vocabulary informants have
with regards to the subject of inquiry. Carley and Palmquist (1992)
report that 29 undergraduates mentioned up to 244 concepts on research
writing and 45 students produced 217 concepts on tutor selection.
One individual could only consider between 30 and 40 concepts in a
session lasting between 20-40 minutes because of combinatorial explosion.
Nakamura et al. (1982) obtained 152 concepts and 265 connections from
5 documents on traffic problems in Japan. Since people have shared
concepts as the number of concepts increase the rate of increase in
total number of concepts will quickly approach zero. The i.i.d. assumption
is reasonable since the stakeholder groups are represented by separate
individuals and they focus on the subject that is only the Kizilirmak
Delta, so the domain focus corresponds to identical distribution (Kosko
1987; Kosko 1988; Taber and Siegel 1987).

\subsection*{Analyzing Cognitive Maps}

Cognitive models are complex systems because they are made of a large
number of units that have many interconnections. This structure results
in an over all behavior of the system that is different than the sum
of units. Therefore, the analysis of complex cognitive maps is difficult.
There are no established statistical methods except for comparing
concept and statement numbers in discrete or continuous categories
with standard statistical techniques (Palmquist et al. 1997). The
greater number of variables (N) in a cognitive map, the more complex
the map is. However, matrix algebra tools of graph theory provide
us with many more indices besides number of variables (concepts, statements)
and their intersections. Comparing number of variables and indices
of cognitive maps across different subject areas and interviewers
might not be correct, because cognitive maps are dependent on the
length of text or duration of interview and the skill of the interviewer
(Eden et al. 1992). But comparisons by standard statistical tests
among stakeholder maps on the same subject area elicited by the same
interviewer as in the case of the Kizilirmak Delta is appropriate.
In this study, Student's t-test was used for comparisons among indices.
If samples were rejected to be normal based on Shapiro-Wilk and Kolmogorov-Smirnov
Tests of Normality, non-parametric Mann-Whitney test was used instead
of t-test.

Graph theory methods help analyzing the structural properties of cognitive
maps, yet another way of analyzing cognitive maps is to use the above
described computational neural network method. The computational method
is a black-box modeling approach which is not necessarily concerned
about the structure, but the outcome of the map. The results of the
cognitive map simulation runs can then be compared, given a fixed
structure of network because \char`\"{}the dynamics of the entire
network must respect this structure\char`\"{} (Reimann 1998).

Another way of analyzing the structure of cognitive maps is to look
how connected or sparse the maps are. There is information contained
in the defined variables as well as the connections between them.
The \textit{density} of a cognitive map ($D$) is an index of connectivity:

\[
D=\frac{C}{N(N-1)}\textrm{ or alternatively, }D=\frac{C}{N^{2}}\]

In the \textit{density} equation $C$, the number of connections,
is divided by the maximum number of connections possible between $N$
variables (Hage and Harary 1983); if variables can have a causal effect
on themselves then maximum number of connections is $N^{2}$.

The structure of a cognitive map apart from number of variables and
connections can best be analyzed by finding \textit{transmitter variables}
(forcing functions, givens, tails), \textit{receiver variables} (utility
variables, ends, heads) and \textit{immediate domains} (Bougon et
al. 1977; Eden et al. 1992; Harary et al. 1965). These variables are
defined by their \textit{outdegree} {[}$od(v_{i})${]} and \textit{indegree}
{[}$id(v_{i})${]}. Outdegree is the row sum of absolute values of
a variable in the adjacency matrix and shows the cumulative strengths
of connections ($a_{ij}$) exiting the variable:

\[
od(v_{i})=\sum_{k-1}^{N}\bar{a_{ik}}\]

\textit{Indegree} is the column sum of absolute values of a variable
and shows the cumulative strength of variables entering the unit.

\[
id(v_{i})=\sum_{k-1}^{N}\bar{a_{ki}}\]

The immediate domain of a variable is the summation of its indegree
(in-arrows) and outdegree (out-arrows) also called centrality. The
contribution of a variable in a cognitive map can be understood by
calculating its \textit{centrality} ($c$) whether it is a transmitter,
receiver or ordinary variable. The centrality ($c$) of a variable
is also called its total degree {[}$td(v_{i})${]} (Harary et al.
1965):

\[
c_{i}=td(v_{i})=od(v_{i})+id(v_{i})\]

In fuzzy cognitive maps, in contrast to binary cognitive maps, a variable
can be more central although it has less connections if the connections
carry larger weights (Kosko 1986).

Transmitter variables are units whose $od(v_{i})$ is positive and
their $id(v_{i})$ is 0. Receiver variables are units whose $od(v_{i})$
is 0 and their $id(v_{i})$ is positive. Other variables which have
both non-zero $od(v_{i})$ and $id(v_{i})$ are ordinary variables
(means) (Bougon et al. 1977).

The total number of receiver variables in a cognitive map can be considered
an index of its complexity. Larger number of receiver variables indicate
that the cognitive map considers many outcomes and implications that
are a result of the system (Eden et al. 1992). On the other hand many
transmitter variables indicate thinking with top down influences,
a \char`\"{}formal hierarchical\char`\"{} system (Simon 1996)
p. 185). Many transmitter units shows the \char`\"{}flatness\char`\"{}
of a cognitive map where causal arguments are not well elaborated
(Eden et al. 1992). Then we can compare cognitive maps in terms of
their complexity by number of \textit{receiver to transmitter variable
ratios} (R/T). Larger ratios will indicate more complex maps, because
they define more utility outcomes and less controlling forcing functions.

Another structural measure of a cognitive maps is the \textit{hierarchy
index} ($h$) (MacDonald 1983):

\[
h=\frac{12}{(N-1)N(N+1)}\sum_{i}[od(v_{i})-(\sum od(v_{i}))/N]^{2}\]

Where $N$ is the total number of variables. When h is equal to one
then the map is fully hierarchical and when $h$ is equal to zero
the system is fully democratic. In a similar schematic Sandel (1996)
calls these domination (hierarchical) and adaptation eco-strategies
(democratic) pointing that democratic maps are much more adaptable
to the local environmental changes because of their high level of
integration and dependence.

\subsection*{Model Abstraction: Simplifying Cognitive Maps by Condensation}

The methods of analysis presented so far have all dealt with characterizing
the whole map with an index for comparison or the contribution of
one variable (centrality) to the map. Although these help us find
important variables and compare different maps using indices, it does
not contribute to a holistic understanding of how the map operates.
Making sense of complex maps is difficult and twenty to thirty variables
start being counterproductive for gaining insights. About a dozen
variables seems to be typical in analysis (Buede and Ferrell 1993).
The best way of approaching an understanding of complex maps is to
simplify them.

How would one go about simplifying a map? According to graph theory
an effective way to better understand the structure of complex cognitive
maps is condensing them. Condensation is achieved by replacing subgraphs
(consisting of a group of variables connected with lines) with a single
unit (Harary et al. 1965). The connections of variables within subgraphs
with other subgraphs are maintained when replacing groups of variables.
The grouping and replacement is also called \textit{aggregation}.
The second question is then how do we decide on subgraphs?

There are two ways of deciding on subgraphs, \textit{qualitative aggregation}
and \textit{quantitative aggregation}. In qualitative aggregation
variables can be combined by categories that are represented by a
larger encompassing variable. Hence Nakamura (Nakamura et al. 1982)
condensed 152 variables coded from 5 documents into 16 categories.
Similarly, in this study I combined 136 categories into 12 overarching
variables based on the categories that have emerged after the listing.
These categories emerge from themes that have been stated by the informants
but also from their relevance to larger theoretical frameworks of
current day scholarship, such as adaptive ecosystem management, community
wildlife management, hydrogeomorphic controls, sustainable livelihoods
and agriculture, ecological economics and common property resource
management. The second way of condensing is quantitative aggregation.
In this case one draws the graphical representation of the cognitive
map and visually defines the strong components (re-enforcing cycles)
as subgraphs (Harary et al. 1965). This approach is similar to the
concept of near-decomposability (Iwasaki and Simon 1994). If we rearrange
the adjacency matrix $A$ so that the sub-adjacency matrices $S_{n}$
of subgraphs align along the diagonal in the form,

\[
A=\left|\begin{array}{cccc}
S_{1} &  &  & \epsilon\\
 & S_{2}\\
 &  & \ddots\\
\epsilon &  &  & S_{N}\end{array}\right|\]

\noindent and all elements outside the submatrices are zero, then
the matrix is considered \textit{completely decomposable}. However
if some small number outside the subgraph is not equal to zero ($\epsilon\not=0$)
then the matrix is nearly decomposable. In a nearly decomposable system
variables within a subsystem interact strongly, but their interaction
to other variables of subsystems is weak. In nearly decomposable systems,
variables in each subsystem move towards a relative equilibrium in
the short-run and the system will run into an over-all equilibrium,
maintaining relative equilibrium in each subsystem on the long-run.
Naturally, the condensed cognitive map based on near-decomposability
will approximate the full map better if $\epsilon$ is smaller (Iwasaki
and Simon 1994). The assumption of near decomposability is valid in
general, because in social and ecological systems it is rare that
every variable is connected to every other variable with equal strength
(Simon 1996).

So how would one decide on submatrices and arrange them along the
diagonal? Cluster analysis can be used to decide on variables that
make up subsystems in a cognitive map. Clustering enables us to find
subsystems that have variables strongly interconnected with each other
and have weak inter-group or variable links (Cossette and Audet 1992;
Eden 1988). Most clustering analysis techniques are not overlapping
(non-inclusive). However in inclusive clustering, recurring variables
in clusters are important because of their ramifications in several
different clusters (Eden et al. 1992). These recurring variables play
central roles in cognitive maps. Cognitive maps that can be readily
clustered or are nearly decomposable are thought to be more simple
compared to ones where clustering is not as easy (Eden et al. 1992).
The hierarchical and k-means clustering of the Kizilirmak Delta social
cognitive maps resulted in one big cluster and several small ones.
The clusters did not correspond to readily recognizable categories.
This shows that the social cognitive maps of the stakeholder groups
are very complex. On the other hand clustering shows hierarchy, and
hierarchy is a property of complex systems (Simon 1996). The issue
of hierarchy was approached with the previously described hierarchy
index, and the failure to detect a meaningful clustering indicates
that the social cognitive maps have little structural complexity based
on Simon's concept of hierarchy. If clusters were achieved they would
have pointed to a hierarchical or stratified system of layers, functional
subsystems, or just divisions depending on the modelers intentions.
We achieve this understanding by the the qualitative aggregation method
described before.

Once the aggregation has been done, the new simplified system can
be represented as a weighted digraph. While drawing the digraph, the
cognitive map connections were drawn so that they reflect the weight
and sign of the causal relationship. I call this representation cognitive
interpretation diagram (CID) similar to the previously developed neural
interpretation diagram (NID) (Ozesmi and Ozesmi 1999).

\section*{RESULTS AND DISCUSSION}

\subsection*{The Content and Structure of Individual Cognitive Maps}

The first step in analyzing cognitive maps is to describe and tabulate
the variables, connections and the structural indices. The description
and the tables can then be used to compare different stakeholder groups.
In the 31 cognitive maps I analyzed on the Kizilirmak Delta, the average
number of variables was 19$\pm$7 SD in the range of 9-34 variables.
The average transmitter variables were 7.0$\pm$4.3 SD, average receiver
variables were 3.0$\pm$2.4 SD, average ordinary variables were 8.9$\pm$3.3
SD per cognitive map. The maps had on average 28.3$\pm$10.6 SD connections
that resulted in a density of 0.112$\pm$0.109 SD. The hierarchy indices
were on average 0.082$\pm$0.135 SD. In a study of a small business
environment, 57 concepts and 87 links were mentioned by the owner
(Cossette and Audet 1992) which results in a connection to variable
ratio of 1.53 and a density of 0.027. A study with a modal average
of 75 minutes of interviewing of 116 informants elicited 32 concepts
in the range from 14-59 (Brown 1992). Eden et al. (1992) found typical
1.15-1.20 ratios of links to nodes. I found 1.64$\pm$0.95 SD ratio
of connections to variables which is comparable to the above mentioned
studies.

The goal of this study is to compare cognitive maps of stakeholder
groups and see how similar and different they are, and then find the
differences and similarities both in structure and functioning in
order to develop strategies that will enable stakeholders to strive
for the conservation of the Kizilirmak Delta. At the start I made
pair wise comparison to understand the association of cognitive maps
of the stakeholders based on which variables they include (Table 1).

\begin{table}[ht]
\begin{center}
\caption{The comparison of similarity and dissimilarity in pairs between cognitive maps of the stakeholders.}
\vspace{0.1in}
\begin{tabular}{lccc}
&McNemar Test&Phi value&Yule Q\\
\hline
Villager - Vacation H. Owner&84.375&0.038&0.146789\\
Villager-NGO Officer&29.762&-0.186&-0.432704\\
Villager-Government Officer&22.413&-0.117&-0.283854\\
Villager-NGO\&Government&7.579&**-0.326&***-0.836158\\
NGO-Vacation H. Owner&29.630&0.146&0.407207\\
NGO-Government Officer&*1.884&**0.370&***0.656566\\
NGO-NGO\&Government&26.000&0.678&1.000000\\
Gov.-Vacation H. Owner&40.695&0.179&0.509982\\
Gov.-NGO\&Government&17.000&0.775&1.000000\\
NGO\&Government-Vacation&60.500&0.173&0.560510\\
\hline
\end{tabular}
\end{center}
\begin{spacing}{1}
\noindent
* The McNemar Symmetry Chi-Square Test for association among matched pairs on shared variables shows that all stakeholder groups are significantly different except between NGO and Government officials.

\noindent
** The phi value indicate the degree of similarity, where 1 is most similar. According to phi values the most similar cognitive maps are those of government and NGO officials maps and the most dissimilar ones are that of villagers compared to government and NGO officials. 

\noindent
*** Yule Q Coefficient is the proportionate reduction in errors in predicting whether or not one group has the variable based on the knowledge that the other group has that variable. The Yule Q value corroborates the results from Phi values.
\end{spacing}
\end{table}

The comparison of the cognitive maps of stakeholders reveals that
stakeholder groups are all significantly different from each other
with the exception of NGO officials and government officials (Table
1). The low phi value and high Yule Q coefficient also indicates that
NGO and government officials are the most similar among stakeholder
groups. There was no significant difference among number of variables
in NGO and Government officials cognitive maps (t-test, p=0.346),
but the power of analysis was only 0.2. The results from McNemar test,
Phi and Yule-Q coefficients supported by t-test prompted me to decide
a priori to pool NGO and Government officials data, and only conduct
statistical tests between villagers and pooled NGO and Government
Officials (Table 1 and 2). Pooling NGO and government data is further
justified by intensive communication between NGO and government officials
and because they were referencing each other during interviews. Similarly
Carley (1986) concludes that the degree of cognitive consensus is
greater in tight social structures with high internal interaction.

The most different stakeholders are Villagers and NGO and government
officials both in terms of phi values and Yule Q coefficient (Table
1). The villagers form the stakeholder group that is most impacted
by any NGO or government project. The significant difference observed
in cognitive maps between villagers and NGO and government officials
separately and pooled indicates that before any project is implemented
there is an important need to base these projects on the understanding
of the villagers (Table 1). The villager-centered approach will avoid
top down imposed projects. A villager-centered approach is not only
necessary because top down projects that do not take into account
local knowledge systems usually fail but because it is the ethical
and responsible way of doing projects.

Villagers in the Kizilirmak Delta had an average of 22.0 variables,
which is significantly more than the pooled NGO and Government Officials'
16.8 average variables (Table 2). The lower number of variables in
NGO and government officials cognitive map compared to villagers does
not necessarily mean that their understanding of the system is less
competent than villagers. Klein and Cooper found that the quality
of decision making was not correlated with the number of variables
(Klein and Cooper 1982). The comparison of number of variables indicate
that villagers have more complex maps than NGO and government officials,
because they have significantly larger number of variables. However
looking at the receiver to transmitter ratio, a better indicator of
structure, we cannot say that they are structurally complex. Transmitter
variables (higher in villagers) indicate a system that is controlled
by forcing functions, a system that is controlled by a formal hierarchy,
not a complex system with embedded \textit{structural hierarchies}.
NGO and government officials maps are also significantly denser with
a larger hierarchy index, which is an indication of structural complexity
(Table 2).

\begin{table}[ht]
\begin{center}
\caption{The average number of different variable types and indices for stakeholder groups in the Kizilirmak Delta.}
\vspace{0.1in}
\begin{tabular}{p{1.4in}lp{0.9in}llp{1in}p{1in}}
&Villagers&Vacation H. \par Owners&NGO Officials&Government \par Officials&NGO \& Govern. Officials\\
\hline
Number of Cognitive \par Maps&15&2&7&7&14\\
Number of Variables \par Mean $\pm$ SD&$^A21.9\pm7.0$&$12.0\pm1.4$&$15.1\pm6.2$&$18.4\pm6.3$&$^A16.8\pm6.3$\\
Number of transmitter \par Variables&$^B9.3\pm3.8$&$5.0\pm0.0$&$3.6\pm3.1$&$6.0\pm4.5$&$^B4.8\pm3.9$\\
Number of Receiver \par Variables&$3.1\pm2.3$&$2.5\pm0.7$&$2.3\pm2.4$&$3.7\pm3.0$&$3\pm2.7$\\
Receiver/Transmitter \par Variable&$^C0.368\pm0.313$&$0.500\pm0.141$&$0.815\pm0.732$&$0.532\pm0.598$&$^C0.830\pm0.688$\\
Number of Ordinary \par Variables&$9.5\pm3.6$&$4.5\pm0.7$&$9.3\pm1.9$&$8.7\pm3.8$&$9\pm2.9$\\
Number of Connections&$28.1\pm9.8$&$14.5\pm5.0$&$28.9\pm9.5$&$32.0\pm12.9$&$30.4\pm11.0$\\
Connection/Variable&$^D1.3\pm0.2$&$1.2\pm0.3$&$2.1\pm1.1$&$2.1\pm1.5$&$^D2.1\pm1.3$\\
Density&$^E0.065\pm0.028$&$0.099\pm0.011$&$0.177\pm0.141$&$0.151\pm0.160$&$^E0.164\pm0.146$\\
Hierarchy&$^F0.029\pm0.027$&$0.075\pm0.027$&$0.149\pm0.130$&$0.131\pm0.241$&$^F0.140\pm0.186$\\
\hline
\end{tabular}
\end{center}
\begin{spacing}{1}
$^A$Number of variables significantly higher in villagers than government and NGO officials together (one-tailed t-test p=0.023).

\noindent
$^B$Number of transmitter variables significantly higher in villagers than government and NGO officials together (one-tailed t-test p=0.002).

\noindent
$^C$Complexity significantly higher in government and NGO officials together than villagers (one-tailed t-test p=0.039).

\noindent
$^D$Connection to variable ratio significantly higher in government and NGO officials together than villagers (one-tailed Mann-Whitney test p=0.025).

\noindent
$^E$Government and NGO officials maps are significantly denser than that of villagers (one-tailed Mann-Whitney test p= 0.015).

\noindent
$^F$Government and NGO officials cognitive maps are significantly more hierarchical compared to villagers (one-tailed Mann-Whitney test p= 0.004).
\end{spacing}
\end{table}

The comparison of indices point that villagers have a broader understanding
of all the variables that affect the Kizilirmak Delta and mention
more variables that control the ecosystem than NGO and Government
Officials. Although there was no significant difference among stakeholder
groups in terms of receiver and ordinary variables, villagers have
a higher number of transmitter variables (Table 2). Larger amount
of transmitters result in a map that has a \textit{formal hierarchy},
yet their cognitive maps are \textit{structurally democratic}, which
indicates that they developed a larger capacity to adapt to changing
conditions. NGO and government officials see fewer variables presumably
because they do not live and earn their living on a daily basis in
the delta. Structurally, they see highly interrelated variables that
are strongly affecting each other and affecting other variables less,
forming \textit{structurally hierarchical} cognitive maps.

This analysis indicates that villagers are faced with many important
forcing functions (transmitter variables) that they can not control.
There is a frustration with these variables and they would like to
see something done about them. Villagers have developed ways of dealing
with these changing and difficult conditions, since their cognitive
maps are highly adaptive, but there is also a message in these maps
to NGO and government officials. Projects which are going to be implemented
in the Kizilirmak Delta have to explicitly address the forcing functions
that impact their daily lives.

The need to address issues of importance to villagers is also supported
by the comparison of 12 most mentioned transmitter variables among
villagers and NGO and government officials. The list reveals the different
concerns by the mentioned forcing functions (Table 3.). There are
only 3 variables that are shared: market forces and middlemen, Bafra
wastewater, and dams. The mentioned variables immediately strikes
us in terms of what are thought to be controlling the ecosystem. For
villagers most of the variables are related to agriculture. Use of
pesticide, fertilizer, government subsidies, drainage, irrigation,
seeds and fuel prices are direct inputs, enhancing the productivity
of agriculture, and are phenomenally all that have made the green
revolution possible. Although these elements have made a strong impression
on the villagers life and thought processes, they see no control or
way of changing these forces. As one villager said: \char`\"{}Maybe
we should all gather and walk to the government center, but no one
has the courage to do so\char`\"{}. The most that they were able
to do was shift, and diversify their economic activities, trying to
adapt to the changing market forces. In the process, however, they
were loosing a considerable amount of income to middlemen. Urban wastewater,
dams industrial effluents and air pollution were seen as external
factors polluting the environment, changing the climate, thereby causing
crop and animal diseases. Water levels were important because high
water limits access to the grazing areas around wetlands. These causal
relationships show that an environmentalism that encompasses these
larger environmental issues as much as biodiversity is more likely
to gain the confidence and cooperation of villagers and might be a
starting point for dialogue. However the biodiversity component should
always be a part of the agenda and must be explained and integrated
to the larger issues as far as the villagers would like to incorporate
it.

\begin{table}[ht]
\begin{center}
\caption{The first twelve most mentioned transmitter variables ranked in order.}
\vspace{0.1in}
\begin{tabular}{lll}
&NGO and Government Officials&Villagers\\
\hline
1.&Incompetence of protection agencies&Use of pesticides\\
2.&NGO's inappropriate policy&Industry\\
3.&County forest protection office&Fertilizer\\
4.&Siltation&Water level\\
5.&Reed collection&Market forces - middlemen\\
6.&Politics and politicial pressures&Bafra wastewater\\
7.&Market forces - middlemen&Dams\\
8.&Law enforcement&Government subsidies\\
9.&Population increase&Drainage\\
10.&Dams&Irrigation\\
11.&Bafra wastewater&Quality seeds\\
12.&Planning and implementation&Fuel prices\\
\hline
\end{tabular}
\end{center}
\end{table}

When we look at the most frequent transmitter variables mentioned
by NGO and government officials we observe that the officials are
very reflexive. NGO and government officials recognize the inadequacy
of protection agencies in realizing their duty and the inappropriate
policies that did not contribute for halting the trend of environmental
degradation and loss of biodiversity in the Kizilirmak Delta. They
think that some of the external factors, that do not have checks and
balances, are causing the degradation of the delta, such as siltation,
politics and political pressures, market forces, the level of law
enforcement, population increase, dams and pollution. They see mostly
reed cutting and planning and implementation as variables that are
positive but could be improved upon. All these mentioned forcing functions
are seen as controlled by an upper hierarchy operating at the national
level. Therefore the NGOs have decided to focus on policy making and
lobbying at the level of the central government with the help of the
sympathetic bureaucrats working in protection agencies. Thereby they
see the solutions to the environmental degradation of the Kizilirmak
Delta, as coming from outside the delta.

\subsection*{The Content and Structure of Social Cognitive Maps}

Social cognitive maps give a larger and more complete picture of the
way in which the stakeholders view the Kizilirmak Delta than individual
cognitive maps. Social cognitive maps are more than a simple addition
of all the separate cognitive maps, because the individual cognitive
maps are already a product of the social cognitive maps as much as
they contribute to it. Therefore an analysis that only looks at individual
cognitive maps is incomplete. The total number of variables defined
in 31 cognitive maps was 136 with 616 connections (Table 4). Out of
136 total variables found in this study, villagers defined 108, vacation
home owners 18, NGO officials 58 and the Government officials defined
67 (See Apendix I). Government and NGO officials together defined
84. To get an idea of the variability in variables and connections
in another subject area, urban development, Malone (1975) reported
that eight graduate students drew a social cognitive map that had
22 variables and 270 connections in a mapping exercise that lasted
two and a half hours.

The indices of the social cognitive maps echo the individual cognitive
maps in some aspects but are different in others (Table 4). Although
we cannot run hypothesis testing statistics on single social cognitive
maps, the differences among stakeholder seems to hold except for the
number of connections and the resulting connection to variable ratio.
The hierarchy index which also takes into account the strength of
connections compared to the number of connections in density is different.
As individual cognitive maps are augmented the possible number of
connections, and the actually drawn connections increase faster than
the number of variables. Therefore the resulting social cognitive
maps are not a simple reflection of the individual maps. The manifestation
of this is the hierarchy index that shows that the social cognitive
map of the villagers and vacation home owners is structurally more
complex than that of NGO and government officials once they are augmented
(Table 4).

\begin{table}[ht]
\begin{center}
\caption{Different variable types and indices for social cognitive maps of the stakeholder groups in the Kizilirmak Delta.}
\vspace{0.1in}
\begin{tabular}{llp{0.8in}lp{1in}p{1in}}
&Villagers&Vacation H. \par Owners&NGO Officials&Government Officials&Total Social \par Cognitive Map\\
\hline
Number of Variables&108&18&58&67&136\\
Number of Connections&382&38&190&204&616\\
Connection to Variable Ratio&3.54&2.11&3.28&3.04&4.53\\
Number of Transmitter Var.&36&6&11&17&27\\	
Number of Receiver Var.&7&4&7&8&9\\	
Receiver to Transmitter Ratio&0.194&0.667&0.636&0.470&0.333\\
Density&0.033&0.117&0.056&0.045&0.033\\
Hierarchy Index&0.030&0.069&0.026&0.021&0.026\\
\hline	
\end{tabular}
\end{center}
\end{table}

Centrality, which is the total indegree reflecting the absolute strength
of connections entering and exiting the variables, is an important
index that reveals which variables play an important central part
in the cognitive maps of the Kizilirmak Delta (Table 5). The similarity
between NGO and government officials is striking. They share 9 of
the most central 12 variables. All stakeholders have put Agriculture,
Livelihood and Fish in a central position. However, NGO and government
officials did not consider animal husbandry to be very central, whereas
the villagers have put it as the most central variable. One of the
major reasons for resistance to protection status from the villagers
is the fear that they will not be able to graze their animals in the
protected zone. Effective protection, political pressures and law
enforcement have been put at a central position by NGO and government
officials and was not considered to be important by the villagers.
In contrast livelihood is put at a higher central position by the
villagers compared to the rest of the stakeholders. Fish are considered
at about the same level of centrality by all stakeholders.

\begin{table}[ht]
\begin{center}
\caption{The most central twelve variables in the social cognitive maps of the stakeholders.}
\vspace{0.1in}
\begin{tabular}{lllll}
&Villagers&Vacation H. Owners&NGO Officials&Government Officials\\
\hline
1.&\textbf{Animal husbandry}&Vacation \& rest&\textbf{Effective protection}&Ecological importance\\
2.&\textit{\underline{\textbf{Agriculture}}}&\textbf{Forest}&\textbf{Political pressures}&\textbf{Effective protection}\\
3.&\textit{\underline{\textbf{Livelihood}}}&\textbf{Wildlife \& birds}&\textit{\underline{\textbf{Agriculture}}}&\textit{\underline{\textbf{Agriculture}}}\\
4.&\textit{\underline{\textbf{Fish}}}&Watermelons \& vegetables&Illegal secondary housing&Pollution\\
5.&Lagoons&\textbf{Animal husbandry}&\textbf{SHW (DSI projects)}&\textbf{Wildlife \& birds}\\
6.&Use of pesticides&The Black Sea&\textit{\underline{\textbf{Fish}}}&Law Enforcement\\
7.&Beach habitat quality&\textit{\underline{\textbf{Fish}}}&\textbf{Law enforcement}&\textit{\underline{\textbf{Fish}}}\\
8.&\textbf{Game wildlife}&\textit{\underline{\textbf{Agriculture}}}&Drainage&\textit{\underline{\textbf{Livelihood}}}\\
9.&Quality grazing area&Angling&\textit{\underline{\textbf{Livelihood}}}&\textbf{Political pressures}\\
10.&Rice agriculture&\textit{\textbf{Livelihood}}&\textbf{Hunters}&\textbf{Hunters}\\
11.&\textbf{Forest}&\textbf{Game wildlife}&NGO  policy&Research \& monitoring\\
12.&Bafra wastewater&Tar \&  petroleum&Water Problems&\textbf{SHW (DSI) projects}\\
\hline
\end{tabular}
\end{center}
\begin{spacing}{1}
\noindent
All variables are ordinary variables except in vacation home owners group "vacation and rest", and "livelihood" are receiver variables, and "tar and petroleum", and "game wildlife" variables are transmitter variables. 

\noindent
\textit{\underline{\textbf{Italic, bold and underlined}}} variables are shared by all stakeholders. \textbf{\textit{Bold and Italic}} variables are shared by NGO and government officials, \textbf{Bold} is shared by two stakeholder groups. Plain font is unique to stakeholder group in the twelve most central variables.
\end{spacing}
\end{table}

The results of Table 5 indicate that NGO and government officials
need to be more sensitive to the issues of animal husbandry, agriculture
and livelihood, especially considering that the other central variables
in the villagers' social cognitive map are directly related to agriculture
and animal husbandry. Effective protection and law enforcement has
to be tied into the concerns of the local people. Wildlife and birds,
fishing and hunting needs to be addressed together with the villagers.
Experiences from community wildlife management projects around the
world can be a starting point for government policies and NGO projects.
Political pressures are part of a democratic system. However if local
support for conservation can be gained, these pressures can be balanced
by local politics and its effect on the larger political arena. Local
support for conservation projects provides a leverage for lobbying
efforts of the NGOs. Currently, the local reaction to conservation
projects only hampers the lobbying efforts of the NGOs despite the
support of government officials. If long term sustainability is the
goal, then conservation can not be achieved without the long-term
support of the villagers. Without the support NGOs and government
officials will constantly be under political pressures that they see
as a central variable in their social cognitive maps.

\subsection*{Condensed Social Cognitive Maps}

The social cognitive maps have a large number of variables and connections
that make it impractical to evaluate them in graphical form. Therefore
the cognitive interpretation diagrams (CIDs) were drawn where 12 variable
categories were defined by qualitative aggregation (Appendix II).
These twelve variables were connected with 36 connections that had
the highest weights. The resulting CIDs (Fig. 4) show that all stakeholders
believe there is a strong negative causal decrease in ecosystem health
caused by environmental problems (i.e., that environmental problems
are serious). NGO and government officials believe that social forces
have an important impact on ecosystem health (Fig. 4b,c). However
while government officials think that social forces can increase ecosystem
health, NGO officials think the social forces impact ecosystem health
negatively and have a virtuous cycle with economic forces. Villagers
believe that ecosystem health causes an increase in resource use and
agricultural production and show an important positive causal relationship
from agricultural production and resource use to livelihood (Fig.
4a). Government and NGO officials do not put this important relationship
in their social cognitive maps (Fig. 4b,c). Water issues are seen
as impacting agricultural production and ecosystem health negatively
by NGOs and government officials. The total social cognitive map of
all stakeholders captures the most important relationships mentioned
in the stakeholder maps and could be used by the parties to come to
a common agreement to form a shared conception of how the Kizilirmak
Delta ecosystem functions. Once that agreement is achieved it can
be a basis for discussions on what the most appropriate policies would
be and which management options to pursue to address the concerns
of all parties.

\subsection*{Cognitive Map Simulations}

Once the cognitive maps are drawn and the adjacency matrix coded,
different simulations can be run and \char`\"{}what-if\char`\"{}
questions can be asked. The behavior of various models can be observed
almost instantly under different scenarios using the neural computational
method. In this study alternative management options were run using
policy, management (independent) and system response (performance,
dependent) variables mentioned by all three stakeholder groups. Shared
variables were chosen so that the results will be most relevant to
all stakeholder groups. The dependent variables were ecological importance,
beach habitat quality, amount of wetland, amount of forest, fish,
wildlife and birds, water problems, pollution, drainage, agriculture,
animal husbandry, logging, and livelihood. These fall into three categories,
ecological importance, environmental problems and livelihood. The
independent variables (policy and management options) were market
forces, law enforcement, municipality, county forestry office, NGO's
inappropriate policies, incompetence of government, effective and
sustained protection (Fig. 5).

In the first no management option all variables were set equal to
one at the start and were allowed to change and settle to a final
value freely. In other options where ecosystem conservation strategies
were simulated independent variables were set (clamped, forced) to
one at each iteration until the system converged. The results of the
social cognitive maps of the stakeholder groups (Fig. 5a,b,c) and
their cumulative cognitive map results (Fig. 5d) indicates that the
best strategy in general is to apply all policy and management options
simultaneously. However the villagers (Fig. 5a) think the best strategy
for beach habitat quality, fish and animal husbandry would be strong
NGO and government policies that would increase their states. But
this strategy would conflict with logging and livelihood, decreasing
their state. The social cognitive map of government officials (Fig.
5c) indicates that they think strong government policies, effective
and sustained protection is going to be a win-win situation increasing
ecosystem importance variables decreasing environmental problems and
increasing livelihood states; villager cognitive maps do not agree.

Two characteristics of cognitive maps becomes visible in these simulations,
the first one is their emergent properties. The sum of the results
of cognitive maps is not the whole, the whole shows different properties
when villagers, NGO and government officials maps are augmented to
form the total cognitive map of all stakeholders (Fig. 5d). The state
of fish and animal grazing is fixed at almost zero under all simulated
options because of the compounded negative causal decrease they receive.
The second characteristic is related and shows the possibility of
synergistic interaction among different management options. For example
the different simulation options do not have an effect on the state
of ecological importance in the cognitive map of all stakeholders,
but when all management options are applied the state will increase
to about 0.2 (Fig. 5d).

An overall look into the three categories of dependent variables indicate
that all stakeholders' cognitive maps agree that the state of ecological
variables (ecological importance, beach habitat quality, amount of
wetland, amount of forest, fish, wildlife and birds) is low compared
to environmental problems (water problems, pollution) and livelihood
(drainage, agriculture, animal husbandary, logging, livelihood) (Fig.
5). However they disagree on how low the level of ecological importance
is. The simulations indicate that villagers believe the ecosystem
is doing better than NGO officials do. NGO officials think it is doing
better than government officials do. However government officials
think it can recover by the various policy and management options
far better than NGO officials think it can (Fig. 5b,c).

The formal validation of these cognitive maps is not possible because
they operate on different understandings of the biophysical and social
spheres of the Kizilirmak Delta. They are also qualitative models
that do not yield outputs measurable in nature. The question whether
some of these maps represent reality better than the others might
not be possible because the reality with which the model outputs are
compared is mediated through yet another understanding. For example
villagers dispute the level of biodiversity assessed by researchers
and say that only commonly breeding birds can be used for evaluating
the importance of the delta for bird species.

\clearpage

\begin{figure}[ht]
\begin{center}
\includegraphics[width=5in]{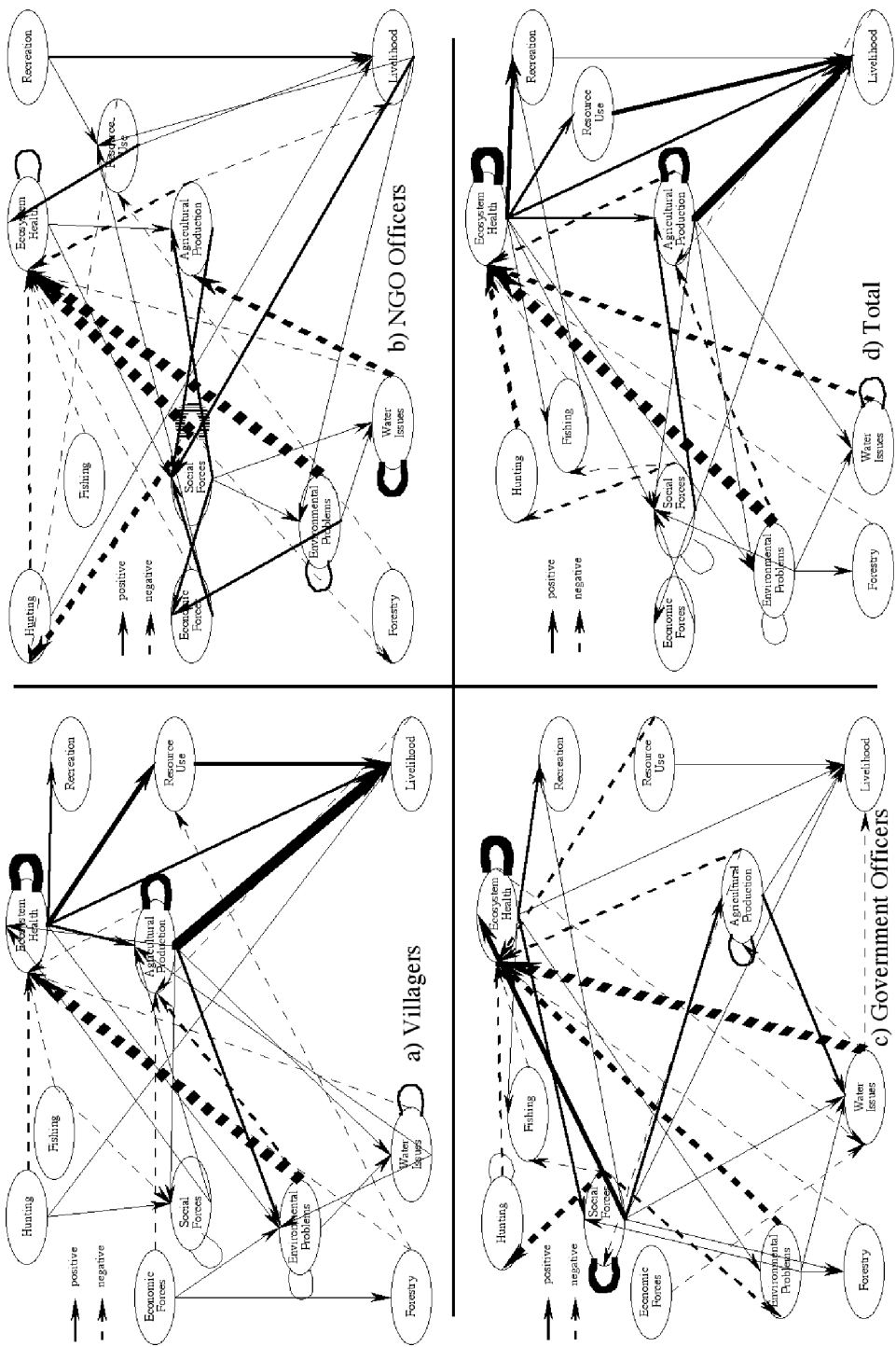}
\caption{Condensed social cognitive interpretation diagrams (CIDs) of a) villagers, b) NGO officials, c) government officials and d) of all stakeholders.  Only the highest 36 connections are shown. Connection arrows have pixel thickness scaled to weights.}
\end{center}
\end{figure}

\clearpage

As what we know and accept as truth is constantly negotiated in society,
the people of the Kizilirmak Delta negotiate what counts as knowledge
and what should be included in the models of the Kizilirmak Delta.
Villagers reflect upon their environment with their own empiricism
and choose and define what counts as knowledge based on their own
normative agenda. On the other hand science that has proven to be
useful for villagers plays an important part in their cognitive maps
as they define variables of the green revolution as transmitter variables
in their maps (Table 3). Nevertheless a qualitative validation can
be made in terms of a \char`\"{}reality check\char`\"{} rather
than formal validation, for example if a map predicts a crash in fish
populations and fishing people are having record harvests year after
year, then obviously something is wrong with the construction of the
model. All stakeholder maps predict that fish stocks are low and villagers
do complain that fishing has gone down in the last 20 years. So some
level of validation can be achieved. An environmental history that
tracks relative changes in variables based both on peoples' perceptions
and historical evidence in records provides an excellent medium to
conduct these reality checks (Ozesmi 1999).

\section*{CONCLUSION}

\subsection*{Utility of Fuzzy Cognitive Maps}

The complexity encountered in the social cognitive maps of the inhabitants
of ecosystems is a humbling experience for the researcher. One is
continuously challenged to describe a thought system that has a complexity
beyond simple representations. Therefore any summary or cross-section
presented is an incomplete picture. A true understanding can only
come through careful exploration of the graphical representation of
the map, looking at summary indices followed by running endless simulations
and observing all the variable changes. This exploration is not reserved
for the researcher, but can be done by the stakeholders as well. The
maps represent an \char`\"{}epistemological structure\char`\"{}
around which stakeholders organize their experience (Cossette and
Audet 1992). The drawing of cognitive maps is a \char`\"{}cathartic
experience\char`\"{} which enables an articulation of how a stakeholder
perceives a system (Eden 1992). The cognitive map in this sense lets
the informant who constructs the cognitive map to explore her present
and future cognition, which facilitates reflectiveness with its \char`\"{}emancipatory\char`\"{}
properties (Cossette and Audet 1992). In addition to the drawing of
the map, the careful exploration and simulation of the individual
and social cognitive maps has to be considered in a continuum both
for the researcher and the stakeholders.

\subsection*{Fuzzy Cognitive Maps for Facilitating Communication}

Sustainability has mostly been formulated by some as resistance movements
against government agendas to develop an ecosystem for economic gain
(Ghai and Vivian 1992). The transformation of the Kizilirmak Delta
into a green revolution \char`\"{}paradise\char`\"{} was accomplished
in the 1950s and 60s before there were any conservation agendas by
NGOs and governments (Ozesmi 1999). In some agricultural systems where
humans see themselves as part of nature with moral obligations, they
have developed conduct that is geared to achieving a balance (Sandell
1996). This is not the case in the Kizilirmak Delta. There is limited
feedback from the wetland ecosystem to economic activities to encourage
ecosystem conservation by villagers. Currently the resistance is,
rightly so, of villagers against sustainable development that restricts
and regulates their economic activity. Yet they also share most of
the concerns with NGOs and government agencies about environmental
degradation. If there is an overlap inter and intra social cognitive
maps of stakeholders, these \char`\"{}may provide a basis for simultaneous
unity and diversity in group processes\char`\"{} {[}emphasis theirs{]}
(Fiol and Huff 1992, p.277). In this way the cognitive maps elicited
in this study are representations of subjective experiences enabling
the stakeholders and the researcher to gain insight on the subject
of study, but more so it is a \char`\"{}tool to facilitate decision-making,
problem-solving, and negotiation\char`\"{} (Eden 1992) to achieve
a normative goal of conservation. If any conservation policies and
ecosystem management is to succeed, it has to take into account the
cognitive maps of the villagers and reconcile it with that of NGOs
and government officials.

\subsection*{Recovery of Local Knowledge, Realism, and Ethics}

The cognitive maps indicate that villagers have significantly larger
number of variables, more complex maps, a broader understanding of
all the variables that affect the Kizilirmak Delta, and mention more
variables that control the ecosystem. They have developed a large
adaptive capacity to changing ecological and social conditions. They
actively change and challenge these conditions through the political
process. Villagers are faced with many important forcing functions
that they cannot control. Most of their variables are related to agriculture
and animal husbandry. This shows that conservation policies and ecosystem
management must encompass these larger environmental issues as much
as biodiversity. One of the major reasons of resistance against a
protection status from the villagers is the fear that they will not
be able to graze their animals in the protected zones. NGOs and government
officials must gain the confidence and cooperation of villagers by
addressing these issues. After this has been accomplished it provides
a starting point for dialogue. Cognitive maps can serve as a basis
when policies and management options are discussed. A villager-centered
cognitive mapping approach is not only necessary because of villagers'
resistance or because top down projects that do not take into account
local knowledge systems fail, but because it is the ethical and responsible
way of doing ecosystem management.

\vspace{-0.2in}

\section*{ACKNOWLEDGEMENTS}

\vspace{-0.1in}

I would like to thank the inhabitants of the Kizilirmak Delta wetlands,
Huseyin and Musa Bas, Fatma and Ahmet Altintas, Isa Eroglu and many
others. Without the valuable information they shared with me this
research would have been impossible. I am particularly grateful to
Sunay Demircan and Sancar Baris for their constant support of this
project. The text benefited immensely from the comments of Herbert
A. Simon, William P. Cunningham, Phillip Regal, David Bengston, Stacy
Ozesmi, and Dale Trexel. This research was funded in part by the MacArthur
Program on Global Change, Sustainability, and Justice, and the Conservation
Biology Program of the University of Minnesota.

\clearpage

\section*{REFERENCES}

\noindent Akkan, E. 1970. Bafra Burnu - Delice Kavsagi Arasindaki
Kizilirmak Vadisinin Jeomorfolojisi (The geomorphology of the Kizilirmak
Valley between Bafra penninsula and Delice joint). Ankara University
School of Language and History - Geography Publications: 191, Ankara.
\medskip{}

\noindent Alkim, U. B., H. Alkim, and O. Bilgi. 1988. Ikiztepe I:
The First and Second Seasons' Excavations (1974-75). Turk Tarih Kurumu
Basimevi, Ankara. \medskip{}

\noindent Alkim, U. B., H. Alkim, and O. Bilgi. 1999. Ikiztepe II:
Excavation Report of Seasons 1976-1980. TÃ¼rk Tarih Kurumu Basimevi,
Ankara. \medskip{}

\noindent Anderson, J. 1983. The Architecture of Cognition. Harvard
University Press, Cambridge, MA. \medskip{}

\noindent Axelrod, R. 1976. Structure of Decision, The Cognitive Maps
of Political Elites. Princeton University Press, Princeton, NJ. \medskip{}

\noindent Bateson, G. 1972. Steps to an Ecology of Mind. Bantam Books,
New York. \medskip{}

\noindent Bateson, G. 1979. Mind and Nature: A Neccessary Unity. Bantam
Boooks, New York. \medskip{}

\noindent Bauer, V. 1975. Simulation, Evaluation and Conflict Analysis
in Urban Planning, Pages 179-192 in M. M. Baldwin, editor. Portraits
of Complexity: Applications of Systems Methodologies to Societal Problems.
Batelle Institute, Columbus, OH. \medskip{}

\noindent Biggs, N. L., E. K. Lloyd, and R. J. Wilson. 1976. Graph
Theory 1736-1936. Clarendon Press, Oxford. \medskip{}

\noindent Bougon, M., K. Weick, and D. Binkhorst. 1977. Cognition
in organizations: an analysis of the Utrecht Jazz Orchestra. Administrative
Science Quarterly 22:606-639. \medskip{}

\noindent Brown, S. M. 1992. Cognitive Mapping and Repertory Grids
for Qualitative Survey Research: Some Comparative Observations. Journal
of Management Studies 29:287-307. \medskip{}

\noindent Buede, D. M., and D. O. Ferrell. 1993. Convergence in Problem
Solving: A Prelude to Quantitative Analysis. IEEE Transactions on
Systems, Man, and Cybernetics 23:746-765. \medskip{}

\noindent Carley, K. 1986. An approach for relating social structure
to cognitive structure. Journal of Mathematical Sociology 12:137-189.
\medskip{}

\noindent Carley, K. 1988. Formalizing Social Expert's Knowledge.
Sociological Methods and Research 17:165-232. \medskip{}

\noindent Carley, K. 1997. Extracting Team Mental Models Through Textual
Analysis. Journal of Organizational Behavior 18:533-558. \medskip{}

\noindent Carley, K., and M. Palmquist. 1992. Extracting, Representing,
and Analyzing Mental Models. Social Forces 70:601-636. \medskip{}

\noindent Coles, J., and B. Coles. 1992. The Wetland Revolution: A
Natural Event, Pages 147-153 in B. Coles, editor. The Wetland Revolution
in Prehistory. The Prehistoric Society, London. \medskip{}

\noindent Cossette, P., and M. Audet. 1992. Mapping of an Idiosyncratic
Schema. Journal of Management Studies 29:325-347. \medskip{}

\noindent Craiger, J. P., R. J. Weiss, D. F. Goodman, and A. A. Butler.
1996. Simulating organizational behavior with fuzzy cognitive maps.
International Journal of Computational Intelligence and Organizations
1:120-133. \medskip{}

\noindent Dickerson, J. A., and B. Kosko. 1994. Virtual worlds as
fuzzy cognitive maps. Presence 3:173-189. \medskip{}

\noindent Eden, C. 1988. Cognitive Mapping. European Journal of Operational
Research 36:1-13. \medskip{}

\noindent Eden, C. 1992. On the Nature of Cognitive Maps. Journal
of Management Studies 29:261-265. \medskip{}

\noindent Eden, C., F. Ackerman, and S. Cropper. 1992. The Analysis
of Cause Maps. Journal of Management Studies 29:309-323. \medskip{}

\noindent Eden, C., S. Jones, and D. Sims. 1979. Thinking in Organizations.
The Macmillan Press, London. \medskip{}

\noindent Fiol, C. M., and A. S. Huff. 1992. Maps for Managers: Where
are we? Where do we go from here? Journal of Management Studies 29:267-285.
\medskip{}

\noindent Ghai, D., and J. M. Vivian. 1992. Grassroots Environmental
Action: Peoples Participation in Sustainable Development. Routledge,
New York. \medskip{}

\noindent Hage, P., and F. Harary. 1983. Structural models in anthropology.
Oxford University Press, New York. \medskip{}

\noindent Hannigan, J. A. 1995. Environmental Sociology: A social
constructionist perspective. Routledge, New York. \medskip{}

\noindent Harary, F., R. Z. Norman, and D. Cartwright. 1965. Structural
Models: An Introduction to the Theory of Directed Graphs. John Wiley
\& Sons, New York, NY. \medskip{}

\noindent Hart, J. A. 1977. Cognitive Maps of Three Latin American
Policy Makers. World Politics 30:115-140. \medskip{}

\noindent Hustings, F., and K. v. Dijk. 1993. Bird Census in the Kizilirmak
Delta, Turkey in Spring 1992. WIWO-report 45, Zeist, The Netherlands.
\medskip{}

\noindent Iwasaki, Y., and H. A. Simon. 1994. Causality and model
abstraction. Artificial Intelligence 67:143-194. \medskip{}

\noindent Jerison, H. J. 1973. The Evolution of the Brain and Intelligence.
Academic Press, New York. \medskip{}

\noindent Kim, H. S., and K. C. Lee. 1998. Fuzzy Implications of Fuzzy
Cognitive Map with Emphasis on Fuzzy Causal Relationships and Fuzzy
Partially Causal Relationship. Fuzzy Sets and Systems 97:303-313.
\medskip{}

\noindent Klein, J. H., and D. F. Cooper. 1982. Cognitive maps of
decision-makers in a complex game. Journal of the Operational Research
Society 33:63-71. \medskip{}

\noindent Kosko, B. 1986. Fuzzy cognitive maps. International Journal
of Man-Machine Studies 1:65-75. \medskip{}

\noindent Kosko, B. 1987. Adaptive Inference in Fuzzy Knowledge Networks,
Pages 261-268, Proceedings of the First IEEE International Conference
on Neural Networks (ICNN-86), San Diego, California. \medskip{}

\noindent Kosko, B. 1988. Hidden Patterns in Combined and Adaptive
Knowledge Networks. Proceedings of the First IEEE International Conference
on Neural Networks (ICNN-86) 2:377-393. \medskip{}

\noindent Kosko, B. 1992a. Fuzzy Associative Memory Systems, Pages
135-162 in A. Kandel, editor. Fuzzy Expert Systems. CRC Press, Boca
Raton, Florida. \medskip{}

\noindent Kosko, B. 1992b. Neural Networks and Fuzzy Systems: A Dynamical
Systems Approach to Machine Intelligence. Prentice-Hall, Englewood
Cliffs, New Jersey. \medskip{}

\noindent Langfield-Smith, K. 1992. Exploring the Need for a Shared
Cognitive Map. Journal of Management Studies 29:349-367. \medskip{}

\noindent Laszlo, E., R. Artigiani, A. Combs, and V. Csanyi. 1996.
Changing Visions, Human Cognitive Maps: Past, Present, and Future.
Praeger, Westport, Connecticut. \medskip{}

\noindent MacDonald, N. 1983. Trees and Networks in Biological Models.
John Wiley and Sons, New York. \medskip{}

\noindent Malone, D. W. 1975. An Introduction to the Application of
Interpretive Structural Modeling, Pages 119-126 in M. M. Baldwin,
editor. Portraits of Complexity: Applications of Systems Methodologies
to Societal Problems. Batelle Institute, Columbus, OH. \medskip{}

\noindent Montazemi, A. R., and D. W. Conrath. 1986. The use of cognitive
mapping for information requirements analysis. MIS Quarterly 10:45-55.
\medskip{}

\noindent Nakamura, K., S. Iwai, and T. Sawaragi. 1982. Decision Support
Using Causation Knowledge Base. IEEE Transactions on Systems, Man,
and Cybernetics SMC-12:765-777. \medskip{}

\noindent Ozesmi, S. L., and U. Ozesmi. 1999. An artificial neural
network approach to spatial habitat modelling with interspecific interaction.
Ecological Modelling 116:15-31. \medskip{}

\noindent Ozesmi, U. 1992. Recent facies in the lagoon barrier east
of the Kizilirmak Delta (Samsun) and ideas on its formation. Geological
Engineering Department, Middle East Technical University, Ankara.
\medskip{}

\noindent Ozesmi, U. 1999. Environmental History Through Faults: Past
and Present Resource Use in the Kizilirmak Delta. unpublished manuscript.
\medskip{}

\noindent Ozesmi, U. 2003. The ecological economics of harvesting
sharp-pointed rush (Juncus acutus) in the Kizilirmak Delta, Turkey.
Human Ecology 31 (4): 645-655. \medskip{}

\noindent Ozesmi, U., and C. Karul. 1990. Bird Observations in the
Kizilirmak Delta, September - 1989, Ankara, Turkey. \medskip{}

\noindent Palmquist, M. E., K. M. Carley, and T. A. Dale. 1997. Applications
of Computer-Aided Text Analysis: Analyzing Literary and Nonliterary
Texts, Pages 171-189 in C. W. Roberts, editor. Text Analysis for The
Social Sciences: Methods for Drawing Statistical Inferences from Texts
and Transcripts. Lawrence Erlbaum Associates, Publishers, Mahwah,
NJ. \medskip{}

\noindent Radomski, P. J., and T. J. Goeman. 1996. Decision Making
and Modeling in Freshwater Sport-fisheries Management. Fisheries 21:14-21.
\medskip{}

\noindent Rappaport, R. A. 1979. Ecology, Meaning, and Religion. North
Atlantic Books, Richmond, CA. \medskip{}

\noindent Reimann, S. 1998. On the design of artificial auto-associative
neural networks. Neural Networks 11:611-621. \medskip{}

\noindent Resmi-Gazete. 1998. Sulak Alanlar Tebligi,, Pages 25-28,
Teblig No:3. Turkiye Cumhuriyeti, Ankara. \medskip{}

\noindent Roberts, F. S. 1973. Building and Analyzing an Energy Demand
Signed Digraph. Environment and Planning 5:199-221. \medskip{}

\noindent Sandell, K. 1996. Sustainability in theory and practice:
a conceptual framework of eco-strategies and a case study of low-resource
agriculture in the dry zone of Sri Lanka, Pages 163-197 in A. Hjort-af-Ornas,
editor. Approaching Nature from Local Communities: Security Percieved
and Achieved. Linkoping University, Linkoping, Sweden. \medskip{}

\noindent Schneider, M., E. Shnaider, A. Kandel, and G. Chew. 1998.
Automatic construction of FCMs. Fuzzy Sets and Systems 93:161-172.
\medskip{}

\noindent Simon, H. A. 1996. The Sciences of the Artificial. The MIT
Press, Cambridge, MA. \medskip{}

\noindent Styblinski, M. A., and B. D. Meyer. 1988. Fuzzy Cognitive
Maps, Signal Flow Graphs, and Qualitative Circuit Analysis, Pages
549-556, Proceedings of the 2nd IEEE International Conference on Neural
Networks (ICNN-87), San Diego, California. \medskip{}

\noindent Taber, W. R. 1991. Knowledge processing with fuzzy cognitive
maps. Expert Systems With Applications 2:83-87. \medskip{}

\noindent Taber, W. R., and M. A. Siegel. 1987. Estimation of expert
weights using fuzzy cognitive maps, Pages 319-325, Proceedings of
the First IEEE International Conference on Neural Networks (ICNN-86).
\medskip{}

\noindent Tekkaya, I., and S. Payne. 1988. The Mammalian Fauna of
Ikiztepe, Pages 227-244 in U. B. Alkim, H. Alkim and O. Bilgi, editors.
Ikiztepe I: The First and Second Seasons' Excavations (1974-75), V.
Dizi Sayi 39 ed. Turk Tarih Kurumu Basimevi, Ankara. \medskip{}

\noindent Tolman, E. C. 1948. Cognitive maps in rats and men. Psychological
Review 55:189-208. \medskip{}

\noindent Wrightson, M. T. 1976. The documentary coding method, Pages
291-332 in R. Axelrod, editor. Structure of Decision, The Cognitive
Maps of Political Elites. Princeton University Press, Princeton, NJ.
\medskip{}

\noindent Zhang, W.-r., and S.-s. Chen. 1988. A logical architecture
for cognitive maps, Pages 231-238, IEEE International Conference on
Neural Networks, San Diego, California. \medskip{}

\clearpage

\section*{Appendix I}

Cognitive map drawn by a villager with corresponding variable translations.

\begin{figure}[ht]
\begin{center}
\includegraphics[width=\textwidth]{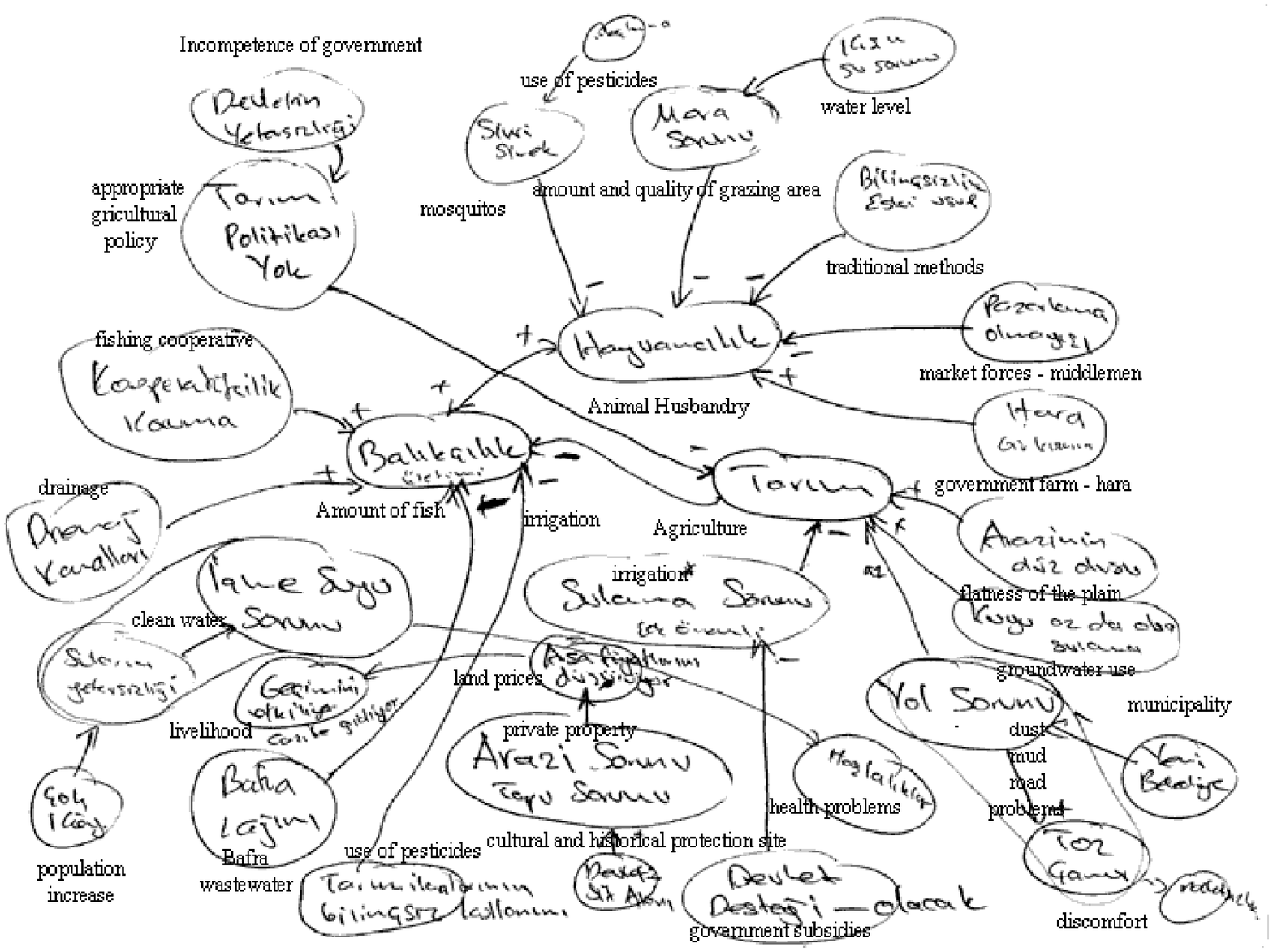}
%\caption{}
\end{center}
\end{figure}

\clearpage

\section*{Appendix II}

Table of all variables and the stakeholder groups that have included
the variables in their cognitive maps, showing whether variables are
transmitter (T), receiver (R) or ordinary (O), and the centrality
of variables scaled between 0 and 1. Variables in bold are shared
at least by villagers, NGO and government officials.

\begin{figure}[ht]
\begin{center}
\includegraphics[height=\textwidth]{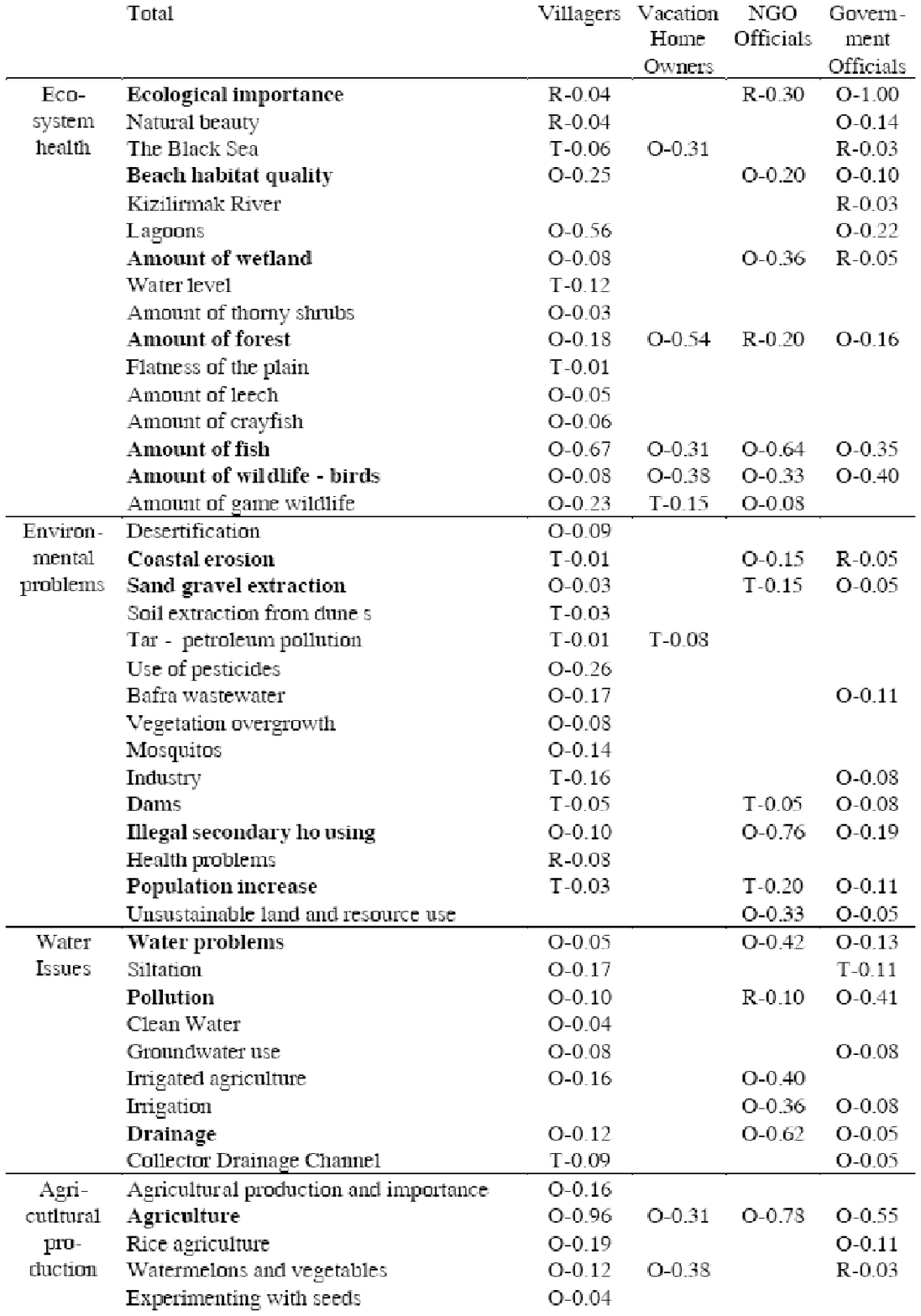}
%\caption{}
\end{center}
\end{figure}

\begin{figure}[ht]
\begin{center}
\includegraphics[height=\textwidth]{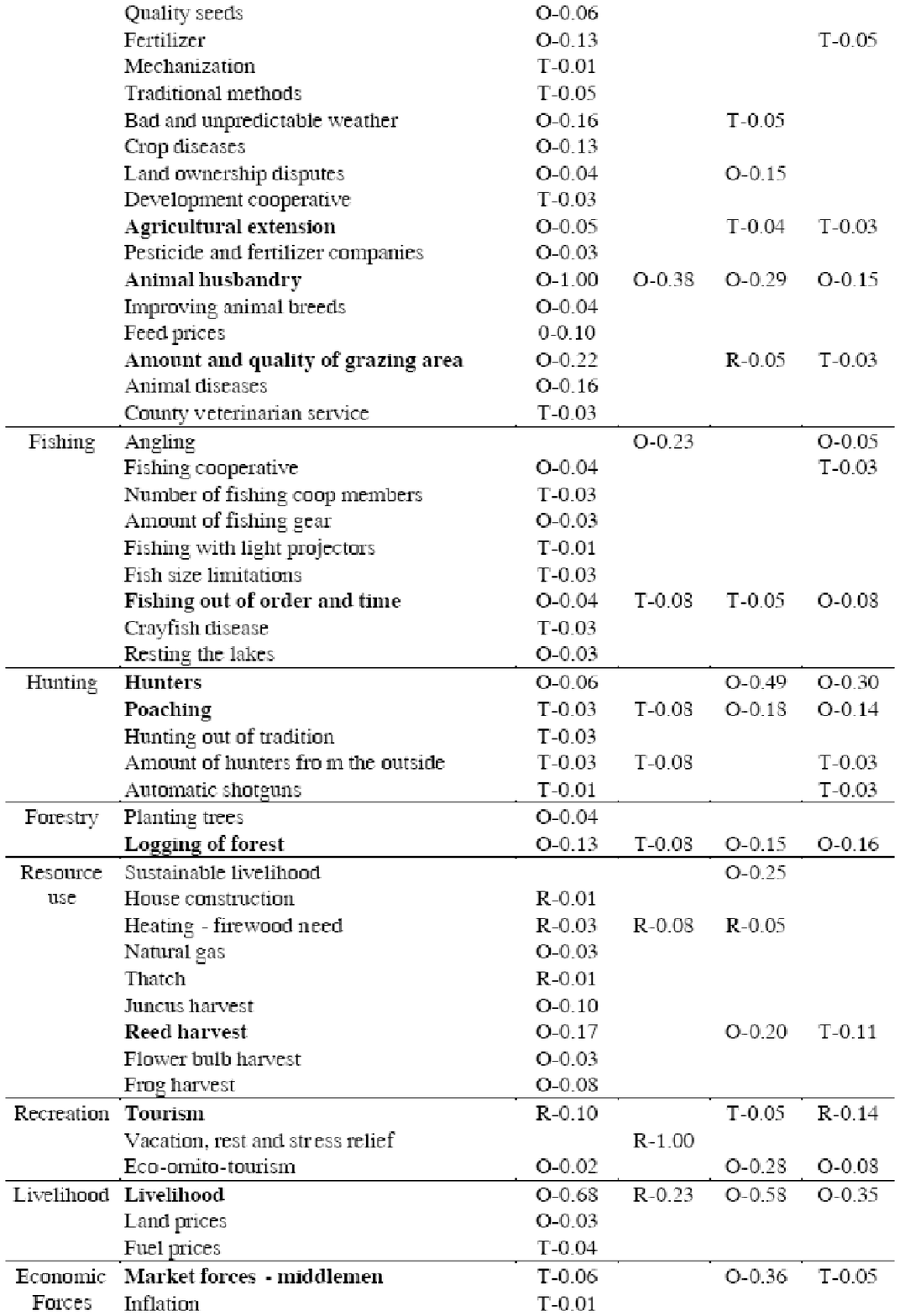}
%\caption{}
\end{center}
\end{figure}

\begin{figure}[ht]
\begin{center}
\includegraphics[height=\textwidth]{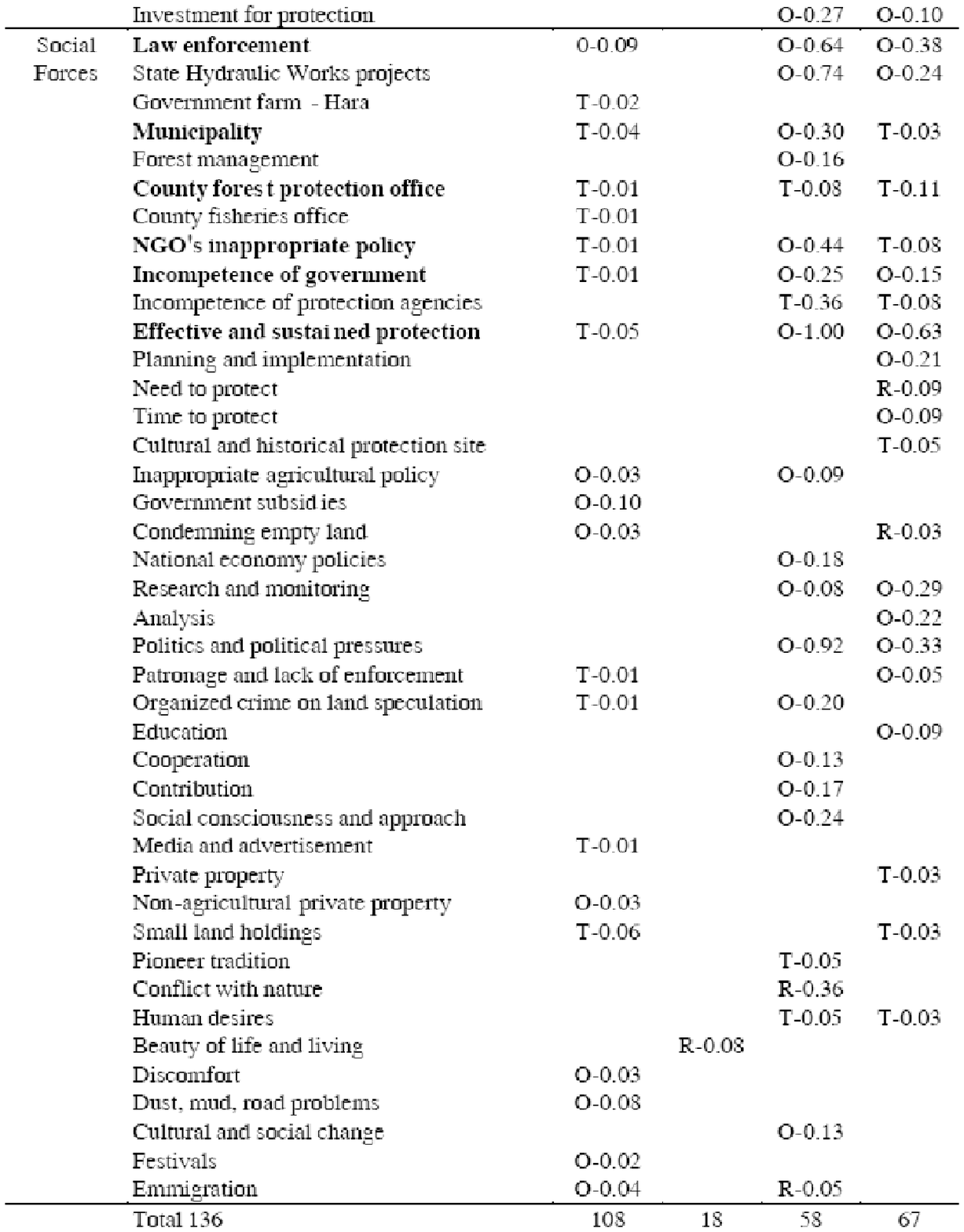}
%\caption{}
\end{center}
\end{figure}

\end{document}